\documentclass[prb,aps,twocolumn,showpacs]{revtex4}
\usepackage{graphicx,color,latexsym}
\usepackage{dcolumn}
\usepackage{amsmath,amssymb,epsf,bm}
\begin{document}
\title{Supercurrent generation by spin injection in an s-wave superconductor-Rashba metal bilayer}
\author{A.~G. Mal'shukov}
\affiliation{Institute of Spectroscopy, Russian Academy of Sciences, Troitsk, Moscow, 108840, Russia}
\begin{abstract}
The spin-galvanic (inverse Edelstein) and inverse spin-Hall effects are calculated for a hybrid system that combines thin superconductor and Rashba-metal layers. These effects are produced by a nonequilibrium spin polarization which is injected into the normal metal layer. This polarization gives rise to an electric potential that relaxes within some characteristic length, which is determined by Andreev reflection. Within this length the dissipative electric current of quasiparticles in the normal layer converts into the supercurrent. This process involves only subgap states and at low temperature the inelastic electron-phonon interactions are not important. It is discussed how such a hybrid system can be integrated into a SQUID  where it produces an effect similar to a magnetic flux.
\end{abstract}
\maketitle

\section{Introduction}

An interplay of the spin-orbit coupling (SOC), magnetism and superconducting correlations in some solids and their interfaces leads to a number of electron transport phenomena, which have attracted recent interest in connection with potential spintronic applications. At the heart of the unusual transport properties of such systems lie the direct and inverse spin-galvanic effects (SGE). The former produces the electric current by polarized spins. The inverse to the SGE effect is also called the Edelstein effect. These phenomena were predicted a long time ago for normal systems. \cite{Ivchenko1978,Levitov,Aronov,EdelsteinSSC} The electric current induced by polarized spins was first observed in a semiconductor quantum well (QW) in Ref.[\onlinecite{Ganichev}], where spin polarization was created by optical excitation. Closely related to SGE are direct and inverse spin-Hall effects (SHE) which convert the charge current into perpendicular spin current and back again. For a review of SHE in normal systems see Ref.[\onlinecite{RashbaSHE}].

There is a fundamental difference between these effects in  superconducting and normal systems. For example, in superconductors the  spin-charge conversion can occur  in  thermodynamic equilibrium conditions. Thus, a spontaneous supercurrent may be produced  by an equilibrium spin polarization induced by a static Zeeman field.  \cite{Edelstein} That is impossible in a  normal metal. This effect, however, can not be observed in spatially uniform systems, because in a weak uniform Zeeman field the so called  helix phase with an inhomogeneous order parameter is formed. \cite{Edelstein,Samokhin,Agterberg1,Kaur,Dimitrova,Agterberg2,Barzykin} In such a superconducting state the electric current is absent. On the other hand, the supercurrent may be induced in the presence of  an inhomogeneous Zeeman field. \cite{Malsh island, Pershoguba,Hals} The equilibrium SGE, as well as the equilibrium analog of the inverse spin-Hall effect, were also predicted in the so called phi-0 Josephson junctions. \cite{Krive,Reinoso,Zazunov,ISHE,Liu,Yokoyama,Konschelle} The supercurrent may also be produced by subgap light illumination of a hybrid superconductor-semiconductor system. \cite{Malsh optics} The inverse SGE was considered for two-dimensional (2D) superconductors and normal metal-superconductor hybrid systems, where the supercurrent gives rise to an equilibrium magnetization by polarizing spins of triplet Cooper pairs.  \cite{Edelstein,Konschelle,Edelstein2,SHE,Bobkov,Bobkov2}

Another group of spin-charge conversion effects involves a nonequilibrium spin polarization, as well as spin current pumped into a superconducting system by some external source. As was shown, for SOC caused by spin-orbit impurities such a nonequilibrium spin distribution can generate the electric current and electric potential in  superconductors. \cite{Takahashi,Espedal}  This nonequilibrium situation resembles much the analogous effects in normal systems. In superconductors, however, the electric and spin transport parameters are determined by quasiparticle characteristics, that are strongly renormalized by the gap in the electron energy spectrum. \cite{Takahashi} In addition, there are typical charge imbalance effects for superconductors, that have not been discussed yet in this context.

In this paper the spin-charge conversion effect will be considered for a bilayer system consisting of a normal metal layer with the strong Rashba SOC and an s-wave superconducting layer. Both layers are coupled through a tunneling barrier. The nonequilibrium spin polarization is injected into the normal layer, as shown in Fig. 1. An advantage of such a system is that it combines strong spin-orbit coupled electrons of the normal metal and correlated Cooper pairs of the superconductor. There are good candidates for the former, such as narrow gap semiconductor quantum wells (QW) and some insulator interfaces. \cite{Lesne,Song} For example, high quality epitaxy grown hybrid semiconductor-superconductor systems have been reported recently. \cite{Shabani,Chang} In its turn, niobium or aluminium films can be employed as
the superconducting layer. The spin polarization can be injected by passing the electric current through a ferromagnetic-paramagnetic interface. \cite{Aronov2,Johnson,Takahashi}

Due to Rashba SOC, in  such a bilayer system the injected spin polarization gives rise to  the electric current inside the normal layer. The mechanisms for such a transformation are SGE and inverse SHE. Note, that in the bounded Rashba systems, such as strips of a finite width, it is impossible to distinguish between these two effects. The electric current
created by these effects is dissipative and is carried by quasiparticles whose energies are below the energy gap of the superconducting layer and above the proximity induced minigap in the spectrum of the normal metal. Then, this current converts into condensate's supercurrent through the Andreev reflection, so that at large enough distances from the point of injection the current is carried by the condensate. The electric potential, that is associated with the quasiparticle's current, also vanishes at large distances together with this current. Such a mechanism of charge imbalance relaxation is the most relevant mechanism in the considered low-temperature regime. It should be noted that in the considered here nonequilibrium system the quasiparticle spins play a major role in the spin-charge conversion. At the same time, in a normal Rashba metal, that contacts to a superconductor, the proximity effect gives rise to triplet Cooper correlations \cite{Edelstein2,Maslov}, so that total spins of correlated electron pairs may potentially contribute to the spin-charge conversion effects. \cite{Bobkov,SHE} The electric current, however, can be produced only if these spins are polarized. In the absence of a Zeeman field they could get some polarization from polarized quasiparticle spins through the electron-electron exchange interaction. \cite{Malsh exchange} This presumably weak interaction will be ignored below, although it can be important in systems with strong exchange effects.

The problem will be considered within the semiclassical approximation. For a dirty system the corresponding  Usadel equations will be employed for the electron Green functions. It is important that, within the main semiclassical approximation, the standard Usadel equations miss the charge-spin coupling, which determines the spin-charge conversion effects. Therefore, such a term will be derived separately, as a linear in $h_F/\mu \ll 1$ nonclassical correction to the Usadel equations, where $h_F$ is the spin-orbit splitting of the electron energy at the Fermi level  $E_F \simeq \mu$ and $\mu$ is the chemical potential.  In this way the expression for the generated electric current, as well as coupled differential equations for the order-parameter phase and quasiparticle distribution function will be obtained. In some important limiting cases these equations will be analyzed analytically.

It is important to emphasize that although the strong SOC favors the spin-charge conversion effects in the normal metal layer, it should not be too strong in the considered case of a disordered system. To reach an injected spin polarization that is high enough, a quasiparticle's spins must survive many collisions with impurities. The corresponding regime of slow D'yakonov-Perel \cite{DP} spin relaxation may be achieved, if the elastic scattering rate is much larger than the spin-orbit splitting of electron energies. This regime can not be realized, for example, for Dirac electrons on  the surface of three dimensional topological insulators, where SOC is comparable to the Fermi energy and the spin relaxation time coincides with the elastic scattering time, because the spin is locked to the electron momentum. For such a material the nonequilibrium SGE could be considered in the clean limit. That is out of the scope of this work.

The paper is organized by the following way. In Sec.II the Usadel equation for the bilayer system will be derived, which phenomenologically accounts for the coupling of the injected spin-dependent distribution to the spin-independent Green function.  In Sec.III such a nonclassical spin-charge coupling term will be calculated in the Usadel equation and in the electric current expression. Also, charge-imbalance relaxation will be analyzed and the differential equation for the phase of the order-parameter obtained. We shall  consider the electric current generated in a closed loop and evaluate the effective electromotive force, that is produced by the spin injection.
\begin{figure}[tp]
\includegraphics[width=8cm]{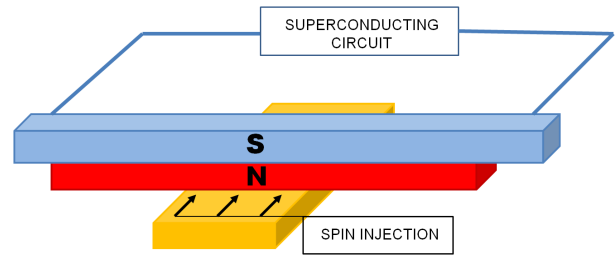}
\caption{(Color online) The bilayer system consists of a Rashba metal "N" and a superconducting layer "S". They are coupled through a tunneling barrier. The spin polarization penetrates into the normal layer from a nonmagnetic lead. In this lead the polarization can be created by spin injection from the ferromagnetic lead (not shown), or otherwise. The superconducting layer may be connected to a superconducting circuit, including e.g. a flux qubit. The spin galvanic effect in the Rashba metal gives rise to the electric current of quasiparticles above the minigap. Due to the Andreev reflection this current transforms in the S-layer into a current of Cooper pairs} \label{fig1}
\end{figure}
\section{Semiclassical equations}

\subsection{Hamiltonian, semiclassical Green functions and self-energies of a bilayer system}

One of the most convenient tools for an analysis of the electron transport  in the range of  characteristic energies $\ll \mu$ and lengths $\gg 1/k_F$, where $k_F$ is the Fermi wave-vector, is a formalism of semiclassical equations for the energy-integrated Green functions.\cite{Eilenberger,Larkin semiclass} This method operates with the three functions $G^r(X_1,X_2)$, $G^a(X_1,X_2)$, and $G^K(X_1,X_2)$, where $X_1=(\mathbf{r}_1,t_1)$  and $X_2=(\mathbf{r}_2,t_2)$  denote space-time variables. $G^r(X_1,X_2)$ and $G^a(X_1,X_2)$ are, respectively, retarded and advanced Green functions, while $G^K(X_1,X_2)$ is the so called Keldysh function. The former carry information about the energy spectrum and wave functions of an electron system, while $G^K(X_1,X_2)$ depends on its statistical properties. It is convenient to combine these functions in the 2$\times2$ matrix $\hat{G}$, such that $G_{11}=G^r$, $G_{22}=G^a$, $G_{12}=G^K$, and $G_{21}=0$. This function satisfies the Dyson equation that can be written in either of two forms, namely, $(i\tau_3\partial_{t_1}-H\tau_3-\hat{\Sigma})\hat{G}=\hat{1}$, or $-i\partial_{t_2}\hat{G}\tau_3-\hat{G}(H\tau_3+\hat{\Sigma})=\hat{1}$, where $H$ is the one-particle Hamiltonian and $\hat{\Sigma}\equiv\hat{\Sigma}(X,X^{\prime})$ is the self-energy matrix. An integration over  intermediate space-time coordinates is implied in the products $\hat{\Sigma}\hat{G}$ and $\hat{G}\hat{\Sigma}$. The Pauli matrices $\tau_1,\tau_2$, and $\tau_3$ operate in the Nambu space. The Hamiltonians of the superconducting and normal layers $H_S$ and $H_N$ have the form
\begin{eqnarray}\label{H}
H_S\tau_3&=&(\epsilon_{S\mathbf{\hat{k}}}-\mu_S)+e\phi_S(\mathbf{r})
-i\texttt{Re}\Delta(\mathbf{r})\tau_2+i\texttt{Im}\Delta(\mathbf{r})\tau_1\nonumber\\
H_N\tau_3&=&(\epsilon_{N\mathbf{\hat{k}}}-\mu_N)+e\phi_N(\mathbf{r})+\mathbf{h}_{\mathbf{\hat{k}}}\bm{\sigma}
\,,
\end{eqnarray}
where $\epsilon_{S\mathbf{\hat{k}}}=\hat{k}^2/2m$, $\epsilon_{N\mathbf{\hat{k}}}=\hat{k}^2/2m^*$, $\mathbf{\hat{k}}=-i\bm{\nabla}_{\mathbf{r}}$. $\mu_S$ and $\mu_N$ are the chemical potentials of electron gases in two layers (the band offset is included). In the case when  the normal layer is  a two-dimensional electron gas, the vector $\mathbf{k}$ in $\epsilon_{N\mathbf{k}}$ has only $x$ and $y$-components, that are parallel to the interface. SOC is represented by the third term in $H_N$, where $\bm{\sigma}=(\sigma_x,\sigma_y,\sigma_z)$ is the vector of Pauli matrices. The spin-orbit field $\mathbf{h}_{\mathbf{k}}=-\mathbf{h}_{\mathbf{-k}}$ is assumed to be a linear function of $\mathbf{k}$. This situation takes place if $\mathbf{h}_{\mathbf{k}}$ is represented by the Rashba field \cite{Rashba} $\mathbf{h}_{\mathbf{k}}=\alpha (\mathbf{e}_z\times\mathbf{k})$, or by the linear Dresselhaus \cite{Dresselhaus} field $h_x=\beta k_x, h_y=-\beta k_y$, as well as by their combination. Below, for simplicity we assume the Rashba SOC. The electric potentials $\phi_S$ and $\phi_N$ appear in Eq.(\ref{H}) due to charge imbalance, that is caused by a conversion of  the injected spin polarization into the electric current of quasiparticles. It should be noted that the spin injection explicitly enters only in the quasiparticle's distribution function, while in  Eq.(\ref{H}) it is represented implicitly through the electric potential and $\Delta$. In the unperturbed state we assume $\texttt{Im}\Delta(\mathbf{r})=0$ and $\texttt{Re}\Delta(\mathbf{r})=\Delta_0$. In principle, the injected spin polarization might enter into Eqs.(\ref{H}) as an effective Zeeman field that is produced by polarized electrons via the electron-electron exchange interaction, as was discussed in Ref.[\onlinecite{Malsh exchange}]. In order to evaluate this field, let us assume that the injected spins are in a quasiequilibrium state that is characterized by the difference $\delta\mu$ between chemical potentials of two spin projections. In this case the effective Zeeman field is $Z\sim G\delta\mu$, where $G$ is the Landau-Fermi liquid exchange parameter. For simple metals $|G|\ll 1$. It is even less in semiconductors, where the Coulomb interaction effects are weaker. In such a case the effect of the Zeeman field may be ignored, because it can not compete with the much stronger effect of quasiparticle spins, that is determined by $\delta\mu$.

In the semiclassical regime $\hat{G}$ varies slowly in both layers as a function of the center of gravity $\mathbf{r}=(\mathbf{r}_1+\mathbf{r}_2)/2$ . At the same time, as a function of $\mathbf{r}_1-\mathbf{r}_2$ it oscillates fast, within the Fermi wavelength. Therefore, it is convenient to Fourier transform $\hat{G}$ with respect to $(\mathbf{r}_1-\mathbf{r}_2)$ and retain  intact its dependence on $\mathbf{r}$. Also, in the considered stationary regime $\hat{G}$  depends only on the time difference $t_1-t_2$ and, hence, can be Fourier transformed to the  frequency variable $\omega$.  Accordingly, let us introduce the Green function as
\begin{equation}\label{Gk}
\hat{G}_{\mathbf{k}}(\mathbf{r},\omega)=\int d^n(\mathbf{r}_1-\mathbf{r}_2)e^{-i\mathbf{k}(\mathbf{r}_1-\mathbf{r}_2)}\hat{G}(\mathbf{r}_1,\mathbf{r}_2,\omega)\,,
\end{equation}
where $n$ is a dimension of the electron gas in a film (the labels N and S are omitted for a while).  The self-energy may be represented in a similar way. The semiclassical Green function $g_{\bm{\nu}}(\omega)$ is defined by integrating Eq.(\ref{Gk}) over  $\xi=\epsilon_{\mathbf{k}}-\mu$ at a fixed direction of $\bm{\nu}=\mathbf{k}/k$ on the Fermi surface. Hence, we have
\begin{equation}\label{gnu}
\hat{g}_{\bm{\nu}}(\mathbf{r},\omega)=\frac{i}{\pi}\int d\xi \hat{G}_{\mathbf{k}}(\mathbf{r},\omega)\,.
\end{equation}
The so defined function is normalized such, that $\hat{g}^2_{\bm{\nu}}=1$. A procedure of obtaining the semiclassical equations for this function is well described in literature. \cite{Larkin semiclass,Rammer, Kopnin} For each layer these, so called, Eilenberger \cite{Eilenberger} equations can be written in the compact form
\begin{equation}\label{Eilenberger}
i\mathbf{v_F}\mathbf{\nabla}\hat{g}_{\bm{\nu}}+[\omega\tau_3 -H\tau_3-\hat{\Sigma}_{\bm{\nu}},\hat{g}_{\bm{\nu}}]=0\,,
\end{equation}
where $\mathbf{v}_F$ is the Fermi velocity.  The nonclassical term associated with the spin-orbit part of the velocity operator $\bm{\nabla}_{\mathbf{k}}(\mathbf{h}_{\mathbf{k}}\bm{\sigma})$ have been neglected in Eq.(\ref{Eilenberger}), because it is small as $h_k/E_F$. It will be included together with other nonclassical terms into a correction to the Usadel equations in Sec.III. The right-hand side of Eq.(\ref{Eilenberger}) should contain the inelastic scattering term. For a considered here situation this scattering is not important. Therefore, it was skipped.

Let us consider  the self-energy term in more detail. First of all, it contains a contribution from electron collisions with impurities.  In the Born approximation for a short-range isotropic scattering amplitude, the corresponding self energy can be written as \cite{Larkin semiclass,Rammer, Kopnin}
\begin{equation}\label{Sigma}
\hat{\Sigma}(\mathbf{r},\omega)=-\frac{i}{2\tau_{\text{sc}}
}\hat{g}(\mathbf{r},\omega)\,,
\end{equation}
where $\tau_{\text{sc}}$ is the elastic scattering time and $\hat{g}(\mathbf{r},\omega)$ is the angular average of $\hat{g}_{\bm{\nu}}(\mathbf{r},\omega)$. Other contributions to the self-energy describe couplings of 2D normal electrons to the superconductor layer and the spin injector. They will be denoted as $\Sigma^{NS}$ and $\Sigma^{NM}$, respectively. Within the semiclassical approach these self-energies are presented only in layers carrying a two-dimensional electron gas, for example,  in a semiconductor QW. At the same time, in a bulk layer, whose thickness is much larger than $k_F^{-1}$, the coupling between layers may be taken into account with the help of boundary conditions for $\hat{g}_{\bm{\nu}}$. The self-energies $\Sigma^{NS}$ and $\Sigma^{NM}$ are determined by virtual electron tunneling from the normal layer to an adjacent layer and back. Let us  assume the normal metal-superconductor tunneling Hamiltonian in the form $\sum_{\mathbf{k},\mathbf{k}^{\prime}} (t^{NS}_{\mathbf{k},\mathbf{k}^{\prime}}\exp[i(\mathbf{k}-\mathbf{k}^{\prime})\mathbf{r}_{NS}]c_{N\mathbf{k}}c^{\dag}_{S\mathbf{k}^{\prime}} +h.c.)$, where $\mathbf{r}_{NS}$ is the interface position in the z-direction. In the $x$ and $y$ directions the interface is homogeneous, so that the parallel wave-vector is conserving. One can easy write the corresponding self-energy in the form
\begin{equation}\label{sigmans}
\hat{\Sigma}^{NS}_{\mathbf{k}}(\mathbf{r})=\sum_{k_z^{\prime},q_z}|t^{NS}_{\mathbf{k},\mathbf{k}^{\prime}}|^2
\int dz \hat{G}_{S\mathbf{k}^{\prime}}(\mathbf{r},z)e^{i\mathbf{q}(\mathbf{z}-\mathbf{z}_{NS})},
\end{equation}
where the vectors $\mathbf{r}$ and $\mathbf{k}$ are  directed along the interface, $\mathbf{q}$ and $\mathbf{z}$ are perpendicular to it, and $\mathbf{k}^{\prime}=\mathbf{k}+\mathbf{k}_z$, with $\mathbf{k}_z$ denoting a vector that is perpendicular to the interface. Within the semiclassical approximation  the $q$-dependence of  $t^{NS}_{\mathbf{k},\mathbf{k}^{\prime}}$ was neglected, because $q$  is small in comparison with $k$ and $k^{\prime}$, which are approximately equal to the electron Fermi wavevector. Then, one may set $k=k_{F_N}$ and $k^{\prime}=k_{F_S}$ in $|t^{NS}_{\mathbf{k},\mathbf{k}^{\prime}}|^2$ , where $k_{F_N}$ and $k_{F_S}$  are the Fermi wavevectors of the normal metal and superconductor, respectively. By integrating $G$ in Eq.(\ref{sigmans})  over energy, we arrive to the  simple expression
\begin{equation}\label{sigmans2}
\hat{\Sigma}_{\bm{\nu}}^{NS}(\mathbf{r},\omega)=-iT_{NS}\hat{g}_{S\bm{\nu^{\prime}}}(\mathbf{r},z=z_{NS},\omega)\,,
\end{equation}
where $T_{NS}=(m/2k_{F_S})|t^{NS}_{\mathbf{k},\mathbf{k}^{\prime}}|_{k=k_{F_N},k^{\prime}=k_{F_S}}^2(\cos\theta_0)^{-1}$ and  the polar angle of $k^{\prime}$ and $\bm{\nu}^{\prime}$ is fixed at $\theta_0$ given by $|\sin\theta_0|=k_{F_N}/k_{F_S}$. The self-energy $\hat{\Sigma}_{\bm{\nu}}^{NM}$, that is associated with a contact to the spin injector, has the  same form as Eq.(\ref{sigmans2}), with $\hat{g}_{S\bm{\nu}}$ and $T_{NS}$ substituted for  $\hat{g}_{M\bm{\nu}}$ and $T_{NM}$. The tunnel coupling with the injector is not zero only in the part of the bilayer system where the normal layer contacts to the injector.

\subsection{Usadel equations for a bilayer system}

Eq.(\ref{Eilenberger}) can be simplified considerably in dirty systems where $v_F\tau_{\text{sc}} \ll v_F/\Delta,v_F/h_{k_F}$ and other  length scales that characterize spatial variations of the Green functions $\hat{g}_{S(N)\bm{\nu}}(\mathbf{r},\omega)$. In this case these functions are almost isotropic and it is possible to obtain  closed equations for their isotropic parts $\hat{g}_{S(N)}(\mathbf{r},\omega)$. \cite{Usadel} The corresponding formalism can be found in reviews Ref.[\onlinecite{Larkin semiclass,Rammer, Kopnin}]. These, so called, Usadel equations in N and S layers can be written in the form
\begin{equation}\label{Usadel1}
D_S\bm{\nabla}\hat{g}_S\bm{\nabla}\hat{g}_S +i[\omega\tau_3+i\bm{\Delta}\bm{\tau},\hat{g}_S]=0
\end{equation}
\begin{equation}\label{Usadel2}
D_N\tilde{\bm{\nabla}}\hat{g}_N\tilde{\bm{\nabla}}\hat{g}_N+i[\omega\tau_3+iT_{NS}\hat{g}_S+iT_{NM}\hat{g}_M,\hat{g}_N]=0 \,,
\end{equation}
where $\bm{\Delta}\bm{\tau}=\texttt{Re}\Delta(\mathbf{r})\tau_2-\texttt{Im}\Delta(\mathbf{r})\tau_1$,$\tilde{\bm{\nabla}}*=\bm{\nabla}*-i[\mathbf{A},*]$ and the gauge-field vector components are $\mathbf{A}^x=-\alpha m \sigma_y,\mathbf{A}^y=\alpha m \sigma_x$ for Rashba SOC. \cite{Konschelle,Gorini,Bergeret} The parameters $D_S$ and $D_N$ denote the electron diffusion coefficients in the superconductor and normal layers, respectively. The first equation is a standard equation of an s-wave superconductor. The second equation contains the spin-orbit effects that are represented by the gauge field $\mathbf{A}$. This equation is written for a 2D electron gas in the normal-metal film. If a 3D gas occupies the film, the  self-energies  $\hat{\Sigma}^{NS}$ and $\hat{\Sigma}^{NM}$ are absent. Instead, a contact with the superconductor and injector can be taken into account with the help of boundary conditions (BC). On the interface between two dirty systems $i$ and $j$, where $i,j=S,N,M$, the conventional form of the boundary condition is \cite{Kupriyanov}
\begin{equation}\label{BC}
D_i\hat{g}_i\nabla_z\hat{g}_i=\gamma_{ij}[\hat{g}_i,\hat{g}_j]\,,
\end{equation}
where the $z$-axis is directed from $i$ to $j$ and $\gamma_{ij}$ can be expressed in terms of the interface resistance. It is expected that Eq.(\ref{BC}) may be modified in the presence of Rashba SOC. In the main semiclassical approximation the modified BC may be obtained in the same way  as in Ref.[\onlinecite{Kupriyanov}]. As a result, $\nabla_z$ is substituted for $\tilde{\nabla}_z$ in the left-hand side of Eq.(\ref{BC}). Both derivatives, however, are equal to each other, because the gauge field $\mathbf{A}_z=0$. The so obtained BC can, in principle, contain nonclassical corrections  $\sim h_{k_F}/E_F$, as was shown for transparent interfaces in clean systems. \cite{Bobkov} In the considered here case of a low transparent interface it is no reason to consider such small terms and the nonclassical corrections will be ignored.

The Usadel equation for $\hat{g}_S$ may be further simplified \cite{Zaitsev} by assuming that Green functions vary slowly  across a thin film, whose thickness $d_S$ is much less than the superconductor's coherence length $\sqrt{D_S/|\Delta|}$. By integrating Eq.(\ref{Usadel1}) over $z$ and taking into account BC Eq.(\ref{BC}) we arrive to the equation for  $\hat{g}_S(\mathbf{r})=(1/d_S)\int dz \hat{g}_S(\mathbf{r},z)$:
\begin{equation}\label{UsadelS}
D_S\bm{\nabla}\hat{g}_S\bm{\nabla}\hat{g}_S +i[\omega\tau_3+i\bm{\Delta}\bm{\tau}+iT_{SN}\hat{g}_N,\hat{g}_S]=0\,,
\end{equation}
where $T_{SN}=D_S\gamma_{SN}/d_S$. If the normal metal is a 3D film, one may perform the same manipulation with a corresponding 3D equation. It should be taken into account that, according to the chosen model, $\tilde{\nabla}_z=\nabla_z$. That results in the equation of the same form as  Eq.(\ref{Usadel2}), with $T_{NS}=D_N\gamma_{NS}/d_N$. The parameters $T_{NS}$ and $T_{SN}$ are related to each other through the equation $d_NN_{F_N}T_{NS}=d_SN_{F_S}T_{SN}$, where $N_{F_N}$ and $N_{F_S}$ are 3D state densities at the Fermi level in the normal metal and superconductor (in the normal state). The same relation will be assumed for $T_{NS}$ entering into the self-energy Eq.(\ref{sigmans2}) of the 2D electron gas, with $d_N N_{F_N}$ substituted for 2D density. Such a relation is necessary for the conservation of the charge current through the NS-interface. It should be noted that the above relation between $T_{NS}$ and $T_{SN}$ means that the latter is much smaller than $T_{NS}$,  when the $N$-film is a semiconductor, but the superconducting layer is a usual metal, whose state density by orders of magnitude exceeds that of the semiconductor.

Further we will focus on some properties of Eqs.(\ref{Usadel2}) and (\ref{UsadelS}) that allow to simplify dramatically the considered problem. First of all, we note that in the case of a spin-injector represented by a massive nonmagnetic film, to which the spin polarization is pumped from ferromagnetic leads, its retarded and advanced Green functions are simply $g_M^{r(a)}=\pm\tau_3$. This means that they are scalars in the spin space. Hence, the retarded and advanced projections of Eq.(\ref{Usadel2}) are scalars, except for terms with the gauge field $\mathbf{A}$. Such terms, however, appear only in the form of commutators of $\mathbf{A}$ with spin scalars $g_N^{r(a)}$. Therefore, they vanish. By representing $\hat{g}_{N(S)}$ in the form $\hat{g}_{N(S)}=\hat{g}_{0N(S)}+\hat{\mathbf{g}}_{N(S)}\bm{\sigma}$ one can see from Eqs.(\ref{Usadel2}) and (\ref{UsadelS}) that $\mathbf{g}_{N}^{r(a)}=\mathbf{g}_S^{r(a)}=0$. Further, let us consider the Keldysh projection of Eqs.(\ref{Usadel2}),(\ref{UsadelS}). In contrast to retarded and advanced functions, the function $g^K_M$ is spin-dependent, because it is determined by the spin dependent distribution function of the injector. Therefore, $g^{0K}_N$, as well as $\mathbf{g}^{K}_N$ are finite. At the same time, one can see from Eqs.(\ref{Usadel2}) and (\ref{UsadelS}) that at $\mathbf{g}^{r(a)}_N=\mathbf{g}^{r(a)}_S=0$ the equations for $g^{0K}_N$ and $\mathbf{g}^{K}_N$ are decoupled from each other. The equations for the three functions $g^K_x,g^K_y$ and $g^K_z$ describe the energy dependent spin diffusion, D'yakonov-Perel spin relaxation and spin precession associated with the Rashba interaction. These processes are essentially the same and described by the same parameters, as in normal metals in the absence of superconducting proximity effects. In other words, the proximity to the superconductor results in the same renormalization factor for all these effects. We will return to the spin transport in Subsection D. The independence of the charge transport on the spin injection is the main reason why one cannot consider the spin-charge conversion effects within the main semiclassical approximation.

\subsection{Retarded and advanced Green functions}

For spin independent retarded and advanced functions ${g}_{0N}^{r(a)}$ one may substitute $\tilde{\nabla}\rightarrow\nabla$ in Eq.(\ref{Usadel2}), so that the spin-charge coupling is only implicitly represented through the phase $\chi(\mathbf{r})$ of the order-parameter $\Delta(\mathbf{r})$. Let us first consider the Usadel equations for the retarded functions by neglecting a contact with the spin injector. We also neglect for a moment the terms containing $\nabla\chi$, that are small, because they are proportional to the weak spin-charge coupling. In such a case ${g}_{0N(S)}^{r(a)}$ do not depend on coordinates. As a result, for the retarded functions Eqs.(\ref{Usadel2}) and (\ref{UsadelS})  reduce  to
\begin{eqnarray}\label{ra}
&&[\omega\tau_3+i\bm{\Delta}\bm{\tau}+iT_{SN}g^r_N,g^r_S]=0\nonumber\\
&&[\omega\tau_3+iT_{NS}g^r_S,g^r_N]=0\,,
\end{eqnarray}
where $\bm{\Delta}\bm{\tau}=\Delta_0[\cos\chi(\mathbf{r})\tau_2+\sin\chi(\mathbf{r})\tau_1]$. A similar equation can be written for the advanced functions. For $T_{NS},T_{SN}$ and $\omega \ll \Delta_0$ the solution of Eq.(\ref{ra}) is given by
\begin{eqnarray}\label{gr}
g^{r(a)}_S&=&\frac{1}{\Delta_0}\left(-i\omega\tau_3+\bm{\Delta}\bm{\tau}+\frac{T_{SN}\omega\tau_3}{\sqrt{(\omega\pm i\delta)^2-\Delta_m^2}}\right)\nonumber \\
g^{r(a)}_N&=&\frac{\omega\tau_3+iT_{NS}(\bm{\Delta}\bm{\tau}/\Delta_0)}{\sqrt{(\omega\pm i\delta)^2-\Delta_m^2}}\,.
\end{eqnarray}
The small terms $\sim T_{SN}/\Delta_0$ and $\omega/\Delta_0$ have been taken into account in $g^{r(a)}_S$, because they are important in the effects associated with the Andreev reflection. One can see that in the quasiparticle spectrum of the normal layer the minigap $\Delta_m=|T_{NS}|\ll \Delta_0$ opens, that is a common property of SN bilayer systems. \cite{Zaitsev}

Since  $\bm{\Delta}$  has a varying in space phase, in some cases it is necessary to take into account corresponding corrections to Eqs.(\ref{gr}). For Green functions of the superconductor these corrections can be easy obtained in the range of high energies $\omega > |\Delta_0|$ by ignoring a contact with the normal layer, whose effect is weak in this frequency range. It is convenient to perform the unitary transformation $g \rightarrow \exp(i\tau_3\chi/2) \tilde{g}\exp(-i\tau_3\chi/2)$. Further, by keeping only linear in $\bm{\nabla}\chi$ terms in Eq.(\ref{UsadelS}) we arrive to the Fourier transformed function $\tilde{g}^{r(a)}_S$ in the form
\begin{equation}\label{gras}
\tilde{g}^{r(a)}_S=\frac{\omega\tau_3 +i\tau_2 \Delta_0}{\Omega}\delta_{\mathbf{q},0} + \frac{\tau_1Dq^2\chi_{\mathbf{q}} \Delta_0}{Dq^2\Omega+2i\Omega^2}
\end{equation}
where $\Omega=\sqrt{(\omega\pm i\delta)^2-\Delta_0^2}$ and $\mathbf{q}$ is the wave-vector. As it will be seen below, this correction is important for calculation of the spin injection effect on the order parameter.

Let us now consider a contact with the injector, as a small correction $\delta g^{r(a)}$ to the functions given by Eqs.(\ref{gr}). Let us  assume that in Fig.1 the length $b$ of the injector-normal metal contact in the $x$-direction is small in comparison with the diffusion length $l_{N}=\left(D_N/2\sqrt{\omega^2-\Delta_m^2}\right)^{1/2}$ ($\omega>|\Delta_m|$). Then, one can represent $T_{NM}(x)$ in Eq.(\ref{Usadel2}) in the form $T_{NM}(x)=bT_{NM}\delta(x)$. For the massive injector film it is also assumed  the unperturbed value $g^{r(a)}_M=\pm \tau_3$. By linearizing Eq.(\ref{Usadel2}) with respect to $\delta g^{r(a)}_N$ we arrive at
\begin{equation}\label{deltag}
\delta g^{r(a)}_N=\mp\frac{bl_{N}}{2l_{NM}^2}\exp\frac{-(1+i)|x|}{\sqrt{2}l_{N}}g^{r(a)}_N\left[g^{r(a)}_N,\tau_3\right]\,,
\end{equation}
where $l_{NM}^2=D_N/T_{NM}$. Therefore, this correction is small at $bl_{N}/l_{NM}^2\ll 1$, that will be assumed in the following.

\subsection{Distribution functions}

In this section we will consider Eqs.(\ref{Usadel2}) and (\ref{UsadelS}) for Keldysh functions. These equations can be transformed to kinetic equations for the distribution function  $f(\mathbf{r},\omega)$, which is defined by the equation \cite{Larkin semiclass,Kopnin}
\begin{equation}\label{f}
g^K=g^rf-fg^a\,.
\end{equation}
The expressed in this way function $g^K$  satisfies the proper normalization condition $g^rg^K+g^Kg^a=0$, that is a nondiagonal projection of the general condition $\hat{g}^2=1$. The distribution function, in turn, can be represented as $f(\mathbf{r},\omega)=f_0(\mathbf{r},\omega)+\mathbf{f}(\mathbf{r},\omega)\bm{\sigma}$. As was noted above, the spin and charge variables are decoupled in Eqs.(\ref{Usadel2}) and (\ref{UsadelS}). This means that we have separate equations for the scalar ($f_0$) and triplet ($\mathbf{f}$) parts of the distribution function.
\subsubsection{Spin distribution function}
Let us first consider the spin-dependent triplet part. We assume that the spin injector is a normal metal, where the spin polarization is creating by electric current passing through a normal metal-ferromagnet interface \cite{Aronov2,Johnson}, or by other means. The thermodynamic equilibrium will be assumed for both spin projections. Hence, the spin distribution function in the injector is given by
\begin{equation}\label{fspin}
\mathbf{f}_M=\frac{\mathbf{s}}{2}\left(\tanh\frac{\omega+\mu_s}{k_BT}-\tanh\frac{\omega-\mu_s}{k_BT}\right)\,,
\end{equation}
where $\mathbf{s}$ denotes the unit vector that is parallel to the spin polarization and $2\mu_s$ is a difference between chemical potentials of spin distributions corresponding to two spin projections. It will be assumed that $\mu_s\ll \Delta_0$. Since $\mathbf{f}_M$ is a scalar in the Nambu space, one can expect that $\mathbf{f}_N$ and $\mathbf{f}_S$ are also scalar functions. Equations for these functions are obtained by substitution of Eqs.(\ref{f}) and (\ref{fspin})
into Eqs.(\ref{Usadel2}) and  (\ref{UsadelS}) and taking the trace over Nambu variables. The terms containing $\nabla \chi$ have been neglected. In this way the equations for  $\mathbf{f}_N$ are obtained in the form
\begin{eqnarray}\label{spinfN}
0=(-\tilde{D}_N\nabla^2+\tilde{\Gamma}_s)\mathbf{f}_{N\parallel}&+&4\alpha m^* \tilde{D}_N\bm{\nabla}f_{Nz}+\nonumber \\
\tilde{T}^{(1)}_{NS}(\mathbf{f}_{N\parallel}-\mathbf{f}_{S\parallel})&+&\tilde{T}^{(1)}_{NM}(\mathbf{f}_{N\parallel}-\mathbf{f}_{M\parallel})\,,\nonumber \\
0=(-\tilde{D}_N\nabla^2+2\tilde{\Gamma}_s)f_{Nz}&-&4\alpha m^*\tilde{ D}_N\bm{\nabla} \mathbf{f}_{N\parallel}+\nonumber \\
\tilde{T}^{(1)}_{NS}(f_{Nz}-f_{Sz})&+&\tilde{T}^{(1)}_{NM}(f_{Nz}-f_{Mz})\,,
\end{eqnarray}
where the labels $\parallel$ and $z$ denote projections of the vector $\mathbf{f}$ onto the $x,y$-plane and the $z$-axis, respectively. Apart from tunneling terms, these equations look as well known spin diffusion equations, \cite{Malsh wire,Mishchenko,Burkov,Malsh accumulation} where spin-charge coupling effects have been neglected.  However, in Eq.(\ref{spinfN}) the spin diffusion  and D'yakonov-Perel spin relaxation coefficients $\tilde{D}_N$ and $\tilde{\Gamma}_s$, respectively, are renormalized by the superconductor proximity effect. The renormalization factor is the same for both transport parameters, such that $\tilde{D}_N/D_N=\tilde{\Gamma}_s/\Gamma_s=(1/4)\mathrm{Tr}[1-g^r_Ng^a_N]$, where $D_N=v_F^2\tau_{\mathrm{sc}}/2$ and $\Gamma_s=2h^2_{k_F}\tau_{\mathrm{sc}}$. The couplings to the superconductor and injector layers are given by
\begin{eqnarray}\label{tildaT}
\tilde{T}^{(1)}_{NS}&=&\frac{T_{NS}}{4}\mathrm{Tr}[(g^r_N-g^a_N)(g^r_S-g^a_S)]\nonumber\\
\tilde{T}^{(1)}_{NM}&=&\frac{T_{NM}}{4}\mathrm{Tr}[(g^r_N-g^a_N)(g^r_M-g^a_M)]\,,
\end{eqnarray}
where $g^r_M-g^a_M=2\tau_3$. It follows from Eqs.(\ref{fspin}-\ref{tildaT}) that the injector spin distribution function plays the role of a source in Eqs.(\ref{spinfN}). At low temperatures the spectral power of this source is distributed in the energy range $\omega\lesssim \mu_s \ll |\Delta|$. At these energies the tunnel coupling between the normal and superconductor layers is weak because $g^r_S-g^a_S$ in $\tilde{T}^{(1)}_{NS}$ is finite only due to subgap quasiparticle states. The contribution of these states is given by the small third term of $g^{r(a)}_S$ in Eq.(\ref{gr}).

In bounded systems Eqs.(\ref{spinfN}) must be appended by boundary conditions. For example, in Fig.1 one needs BC at the edges $y=\pm w/2$, where $w$ is the width of the bilayer. A generalization of BC that takes into account  Rashba SOC has been discussed in Sec.II B. It can be achieved by the substitution  $\nabla_y \rightarrow \tilde{\nabla}_y$ in Eq.(\ref{BC}).  By setting at the edges $\gamma=0$ and multiplying Eq.(\ref{BC}) by $\hat{g_i}$ we get $\tilde{\bm{\nabla}}_y \hat{g}_i=0$. Since the $y$-independent retarded and advanced functions are scalars in the spin space, the latter equation may be reduced, with the help of Eq.(\ref{f}), to a set of  equations for vector components of the spin distribution function. These equations have the form
\begin{equation}\label{bc}
\nabla_yf_z+2\alpha m^* f_y=\nabla_yf_y-2\alpha m^* f_z=\nabla_yf_x=0\,,
\end{equation}
where all functions are taken at $y=\pm w/2$. The same equations take place for the spin density in normal systems, if the diffusive spin current parallel to the $y$-axis turns to zero at the hard wall boundary. \cite{Adagideli,Bleibaum,Brataas NJF,Malsh dipole} Therefore, Eqs.(\ref{bc}) for the distribution function seem reasonable, although this issue deserves a separate study.

Due to spin precession in the Rashba field the boundary conditions Eq.(\ref{bc}) always mix various spin components. For example, if the injected spin polarization  in Eq.(\ref{fspin}) is initially oriented parallel to $y$-axis and homogeneous in the $y$-direction, it will rotate towards the $z$-axis during propagation along the strip. Therefore, it is impossible to observe a pure spin-galvanic effect in bounded systems, because in the  presence of $z$-polarized spins the  inverse spin-Hall effect also takes place.  When the strip width is much larger than the spin diffusion/precession length $l_{so}=1/\alpha m^*$, one may neglect the boundary conditions. In this case  the solution of  Eq.(\ref{spinfN}) is $f_{Nx}=f_{Nz}=0$ and $f_{Ny}$ depends only on the $x$-coordinate, if $\mathbf{f}_M$ is chosen in the form of Eq.(\ref{fspin}) with $\mathbf{s}$ parallel to the $y$-axis. By assuming in Eq.(\ref{spinfN})  $2T^{(1)}_{NM}(x)=T^{(1)}_{NM}[\theta (x+b/2)+\theta (x-b/2)]$, where $\theta (x)$ is the step function, this solution can be obtained  in the form
\begin{eqnarray}\label{fyN}
f_{Ny}&=&A\theta\left(\frac{b}{2}-|x|\right)\left(1-B\cosh2 x\kappa^{\prime}\right)+\nonumber\\
&&AB\theta\left(|x|-\frac{b}{2}\right)\frac{\kappa^{\prime}}{\kappa}\sinh b\kappa^{\prime}e^{-2 x\kappa}
\end{eqnarray}
where $A=\tilde{T}^{(1)}_{NM}f_{My}/(\tilde{\Gamma}_s+\tilde{T}^{(1)}_{NM})$, $B=\kappa(\kappa\cosh b\kappa^{\prime}+\kappa^{\prime}\sinh b\kappa^{\prime})^{-1}$, $\kappa=1/l_{so}$ and $4\kappa^{\prime2}=4\kappa^2+(\tilde{T}^{(1)}_{NM}/\tilde{D}_N)$. We neglected in Eq.(\ref{fyN}) a leakage of the spin polarization into superconductor. Also, it was assumed that $\tilde{\Gamma}_s \gg \tilde{T}^{(1)}_{NS}$. The opposite case of a narrow strip with $w\ll l_{so}$ is  considered in Sec.IIIB2.

\subsubsection{Particle distribution function}
In Eq.(\ref{f}) the spin-independent scalar function $f_0$ can be represented in the form of  a diagonal matrix in the Nambu space.\cite{Kopnin} Accordingly, we have $f_0=f^{(1)}_0+\tau_3f_0^{(2)}$. Let us first consider the Usadel equation for $f_0^{(1)}$. It is important that the spin injector pumps into the system not only nonequilibrium spins, but also particles that are out of the thermodynamic equilibrium. Indeed, the spin-independent distribution function in the injector is
\begin{equation}\label{f0}
f^{(1)}_{0M}=\frac{1}{2}\left(\tanh\frac{\omega+\mu_s}{k_BT}+\tanh\frac{\omega-\mu_s}{k_BT}\right)\,.
\end{equation}
 This function differs from the equilibrium distribution. Such
sort of a distribution function was considered in Ref.[\onlinecite{Takahashi2}]. Let us look what happens in the bilayer geometry shown in Fig. 1. From Eqs.(\ref{Usadel2}) and  (\ref{UsadelS})  the Usadel equations for $f^{(1)}_{0N}$ and $f^{(1)}_{0S}$ can be obtained in the form
\begin{eqnarray}\label{f1NS}
0&=&-\tilde{D}_N\nabla^2f^{(1)}_{0N}+\tilde{T}^{(1)}_{NS}(f^{(1)}_{0N}-f^{(1)}_{0S})+\nonumber \\
&&\tilde{T}^{(1)}_{NM}(f^{(1)}_{0N}-f^{(1)}_{0M})\,,\nonumber \\
0&=&-\tilde{D}_S\nabla^2f^{(1)}_{0S}+\tilde{T}^{(1)}_{SN}(f^{(1)}_{0S}-f^{(1)}_{0N})\,.
\end{eqnarray}
The renormalization factor for  $\tilde{T}^{(1)}_{SN}$ is the same as for $\tilde{T}^{(1)}_{NS}$ in Eqs.(\ref{tildaT}). The diffusion constant  in the superconductor is given by  $\tilde{D}_S=(D_S/4)\mathrm{Tr}[1-g^r_Sg^a_S]$. The solution of Eqs.(\ref{f1NS}) is $f^{(1)}_{0S}=f^{(1)}_{0N}=f^{(1)}_{0M}$. This solution is valid as long as the inelastic scattering was ignored. If inelastic relaxation processes are taken into account, the distribution functions in both layers will relax to the thermal equilibrium at a large distance from the injection point. We will assume that the temperature is small enough, such that this distance is much larger than the spin-orbit relaxation/precession length $l_{so}$ and other characteristic lengths that determine the spin-charge conversion. Since the quasiparticle's energy distribution in the superconductor's layer is different from the equilibrium one, the gap will slightly decrease.  The distribution function in the form of Eq.(\ref{f0}) produces a weak effect at  $k_BT \ll \Delta$ and $\mu_s \ll \Delta$. \cite{Takahashi2} We will assume that this effect is included into the gap. Such a gap depends slightly on $x$ and relaxes together with $f^{(1)}_{0S}$ to its unperturbed value at large distances.

The functions $f^{(2)}_{0S}$ and $f^{(2)}_{0N}$ control the kinetics of the spin-charge conversion. Within the considered model such a function is zero in the injector, i.e. $f^{(2)}_{0M}=0$. Therefore, according to Eqs.(\ref{Usadel1}) and (\ref{Usadel2}), $f^{(2)}$ must be zero in the entire system. On the other hand, these equations miss the spin-charge coupling terms that are responsible for the direct and inverse spin-Hall and spin-galvanic effects. These terms play the role of nondiagonal elements that couple two sets of Usadel equations for spin independent and spin dependent Green functions, $g_0$ and $\mathbf{g}$, respectively. As it will be shown in the next section, in the transport equations for the spin-independent function $g_{0N}^K$ the spin-charge coupling appears as a  proportional to $\mathbf{g}^K$ term. The latter, in turn, can be expressed through the spin distribution function $\mathbf{f}$, that was considered in Subsubsection 1. As a result, the effective $\omega$-dependent  electromotive force $\bm{\mathcal{E}}$, that is given by Eq.(\ref{E}), appears in the equation for $f^{(2)}_{0N}$. By substituting Eq.(\ref{f}) into the scalar projection of Eq.(\ref{Usadel2}) and adding there the spin-charge coupling $\mathcal{E}$ the transport equations for $f^{(2)}$ take the form
\begin{eqnarray}\label{f2NS}
0&=&-\tilde{D}^{(2)}_N\nabla^2f^{(2)}_{0N}-\mathbf{j}_N\bm{\nabla}f^{(1)}_{0N}-eD_N\bm{\nabla}\bm{\mathcal{E}}+\nonumber \\
&&\tilde{T}^{(2)}_{NS}(f^{(2)}_{0N}-f^{(2)}_{0S})+\tilde{T}^{(2)}_{NM}f^{(2)}_{0N}\,,\nonumber \\
0&=&-\tilde{D}^{(2)}_S\nabla^2f^{(2)}_{0S}-\mathbf{j}_S\bm{\nabla}f^{(1)}_{0S}-Rf^{(2)}_{0S}+ \nonumber \\
&&\tilde{T}^{(2)}_{SN}(f^{(2)}_{0S}-f^{(2)}_{0N})\,.
\end{eqnarray}
where the energy dependent transport parameters are \cite{Zaikin rev}
\begin{eqnarray}\label{D2}
&&\mathbf{j}_{N(S)}=\frac{D_{N(S)}}{4}\mathrm{Tr}\left[(g^r_{N(S)}\bm{\nabla}g^r_{N(S)}-g^a_{N(S)}\bm{\nabla}g^a_{N(S)})\tau_3\right]\,,\nonumber\\
&&\tilde{D}_{N(S)}^{(2)}=\frac{D_{N(S)}}{4}\mathrm{Tr}\left[1-\tau_3g^r_{N(S)}\tau_3g^a_{N(S)}\right]\,\, \, \text{and} \nonumber\\
&&R=\frac{D_S}{4}\mathrm{Tr}\left[(g^r_S+g^a_S)\bm{\Delta}\bm{\tau}\right]\,.
\end{eqnarray}
The tunneling parameters are given by
\begin{equation}\label{tildaT2}
\tilde{T}^{(2)}_{AB}=(T_{AB}/4)\mathrm{Tr}[(\tau_3g^r_A-g^a_A\tau_3)(g^r_B\tau_3-\tau_3g^a_B)]\,,
\end{equation}
where $A$ and $B$ take the values $N$, $S$ or $M$. In Eqs.(\ref{f2NS}) the terms with the spectral supercurrents $\mathbf{j}_{N(S)}$ are small because $\bm{\nabla}f^{(1)}_{0N(S)}$ are inversely proportional to the large inelastic relaxation length, as it follows from the above analysis. The spectral supercurrents $\mathbf{j}_{N(S)}$ are also small, because they are proportional to $\nabla \chi$. The latter is determined by the weak spin-charge coupling. Therefore, these terms will be neglected. Furthermore, the coupling to the injector may also be neglected, because the function $f^{(2)}_{0N}$ changes its sign in the range of the injector, that makes very inefficient a leakage of $f^{(2)}_{0N}$ into the injector at $b \ll l_{NM}$.

The transport parameters in Eqs.(\ref{f2NS}) can be calculated by using Eqs.(\ref{gr}) and (\ref{D2}). By taking into account only the leading terms at $\omega \ll |\Delta_0|$ we obtain
\begin{eqnarray}\label{D3}
&&\tilde{D}_{N}^{(2)}=\frac{D_N \omega^2}{\omega^2-|\Delta_m|^2}\,\,, \,\,\tilde{D}_{S}^{(2)}=D_S \,\,, \,\, R=2\Delta_0\exp i\chi  \,,                      \nonumber\\
&&\frac{\tilde{T}^{(2)}_{NS}}{T_{NS}}=\frac{\tilde{T}^{(2)}_{SN}}{T_{SN}}=\frac{2T_{SN} \omega^2}{\Delta_0(\omega^2-|\Delta_m|^2)}\,.
\end{eqnarray}
With these parameters a kinetics of the charge imbalance relaxation, that is controlled by  Eq.(\ref{f2NS}), becomes clear. Indeed, in the normal metal layer the electromotive force $\mathbf{\mathcal{E}}$ generates the nonzero $f^{(2)}_{0N}$. The latter, in turn, is related to the electric potential according to the equation \cite{Kopnin}
\begin{equation}\label{phi}
e\phi=-\frac{1}{8}\int d\omega \mathrm{Tr}[\tau_3(g^r_N-g^a_N)]f^{(2)}_{0N}\,.
\end{equation}
This potential implies a presence of a charge imbalance, that relaxes through tunneling of electrons into the adjacent superconducting layer, where $f^{(2)}_{0S}$ turns to zero relatively fast due to quasiparticle's absorption by the condensate. According to Eqs.(\ref{f2NS}) and (\ref{D3}), the characteristic relaxation length of $f^{(2)}$ in the superconductor is the smallest one of  $l_R$ and $l_{SN}$, which are given by
\begin{eqnarray}\label{lR}
&&l_{SN}=\sqrt{\frac{\tilde{D}^{(2)}_S}{|\tilde{T}^{(2)}_{SN}|}}=\frac{1}{\sqrt{2}\omega |T_{SN}|}\sqrt{|\Delta_0|D_S(\omega^2-\Delta^2_m)}\,,\nonumber\\
&&l_R=\sqrt{\frac{\tilde{D}^{(2)}_S}{|R|}}=\sqrt{\frac{D_S}{2\Delta_0}}\,.
\end{eqnarray}
It is seen from this equation that, except for a narrow region of energies close to $\Delta_m$, the distribution function  $f^{(2)}_{0S}$ vanishes within  the  superconductor's coherence length $l_R$. At the same time, $f^{(2)}_{0N}$ decreases in space only due to a slow leak into the superconductor, within the length $l_{NS}=(\tilde{D}^{(2)}_N/\tilde{T}^{(2)}_{NS})^{1/2} =(D_N \Delta_0/ T_{SN}T_{NS})^{1/2}$, which is assumed much larger than $l_R$. Therefore, $l_{NS}$ determines the charge imbalance relaxation in the whole system. This mechanism  is different from the well known charge imbalance relaxation at superconductor-normal metal interfaces, that involves inelastic electron-phonon scattering.\cite{Artemenko,Tinkham,Schmid}  A relaxation of the electric potential is accompanied by a transformation of the electric current of quasiparticles into the supercurrent. This issue will be discussed in more detail in Sec. IIIB.
\section{Calculation of the spin-charge coupling term in Usadel equations}
\subsection{Quantum corrections to the Usadel equations and electric current}

As was shown in the previous section, the Usadel equations, that were obtained within the main semiclassical approximation, are decoupled into two independent sets of equations for spin- singlet and spin-triplet Green functions. In this section  the quantum correction which leads to a mixing of these two sets will be calculated in the first order with respect to $\alpha/v_F \sim h_{k_F}/\mu$. It follows from Sec.II that, as long as the Fermi liquid effects are ignored, the retarded and advanced functions stay scalar in the spin space. Therefore, let us focus on the Keldysh function. We start from the Dyson equation
\begin{equation}\label{Dyson}
(\omega\tau_3-H_N\tau_3)G^K_{N}=\Sigma^r_N\circ G^K_{N}+\Sigma^K_N\circ G^a_{N}\,.
\end{equation}
In this equation the Green function and self-energy depend on two spatial coordinates and "$\circ$" denotes the integration over an intermediate coordinate. Eq.(\ref{Dyson}) can be simplified by taking into account that the self-energies Eq.(\ref{Sigma}) and (\ref{sigmans}) are local functions of $\mathbf{r}$ and can be expressed through the semiclassical angular averaged Green functions. Also, one should take into account that $G^{r(a)}_{N}$ depend weakly on coordinates, as was discussed in Sec.II. By combining the first term in the right-hand side of Eq.(\ref{Dyson}) with the expression in the left one may express the perturbed part of the Keldysh function in the form $G^K=G^r\circ\Sigma^K\circ G^a$. Further, by transforming $G^K$ to the mixed representation Eq.(\ref{Gk}) and integrating it over $\mathbf{k}$ the Fourier transformed spin-independent part of $g^K(\mathbf{q})$ can be expressed as
\begin{equation}\label{Dyson2}
g^K_{0N}(\mathbf{q})=\frac{1}{2}\sum_{\mathbf{k}} \mathrm{Tr}_{\sigma}[G^r_{N\mathbf{k}+\frac{\mathbf{q}}{2}}\Sigma^K_N(\mathbf{q}) G^a_{N\mathbf{k}+\frac{\mathbf{q}}{2}}]\,.
\end{equation}
One may obtain Usadel equation Eq.(\ref{Usadel2}) for the Keldysh function by expanding $G^{r(a)}_{N\mathbf{k}\pm\frac{\mathbf{q}}{2}}$ in $\mathbf{q}$, $\omega$ and $h_{\mathbf{k}}$ and performing the integration over $\mathbf{k}$ within the main semiclassical approximation. That means that all slowly varying entries in the integral are evaluated at $k=k_F$. Since we are interested in quantum corrections, such terms have to be  expanded near $k_F$. The task, however, is not so vast, because the goal is to calculate only the terms that couple the spin-independent and spin-dependent functions $g^K_{0N}$ and $\mathbf{g}^K_{N}$. Therefore, only the spin-dependent part of $\Sigma^K_N(\mathbf{q})$ will be taken into account in Eq.(\ref{Dyson2}). Furthermore, contributions to $\Sigma^K_N(\mathbf{q})$, that are caused by electron's tunnelings to the injector and superconductor, are much smaller than the self-energy associated with the elastic impurity scattering. Therefore, they will be  ignored below.

At the small frequency $\omega\ll \Delta_0$ the retarded and advanced functions are obtained from  Eqs.(\ref{H}) and  (\ref{Sigma})  in the form
\begin{equation}\label{G}
G^{r(a)}_{N\mathbf{k}}=\frac{1}{4}\sum_{\sigma=\pm 1}\left[\frac{g^{r(a)}_N+1}{\lambda\Omega - {\xi_{\sigma} }}+\frac{g^{r(a)}_N-1}{\lambda\Omega+ {\xi_{\sigma} }}\right](1+\sigma(\mathbf{n}\bm{\sigma}))
\end{equation}
where $\xi_{\sigma}=\xi+\sigma h_{\mathbf{k}}$, $\mathbf{n}=\mathbf{h}_{\mathbf{k}}/h_{\mathbf{k}}$, $\Omega=\sqrt{(\omega\pm i\delta)^2-(\Delta_m)^2}$ and the factor $\lambda$ is given by $\lambda\Omega=\Omega+i\omega/2\tau_{\text{sc}}|\omega|$. The functions $g^{r(a)}_N$ in this equation are given by Eq.(\ref{gr}), where the phase factor in the order parameter is ignored, so that  $\bm{\Delta}\bm{\tau}=\Delta_0\tau_2$. This simplification is dictated by the accuracy of nonclassical corrections, that must be  linear in $h_{k_F}/\mu$. Since the phase is small on this parameter, one must neglect it in $g^{r(a)}_N$. For the same reason, the electric potential $\phi_N$ should also be neglected. In view of the relatively large scattering rate $1/\tau_{\text{sc}}$, the calculation of the integral over $\xi$ in Eq.(\ref{Dyson2}) can be performed by expanding the denominators in Eq.(\ref{G})  with respect to $h_{k}$, $\Omega$ and $q$. Also,  $h_{\mathbf{k}\pm\mathbf{q}/2}$, $\mathbf{n}_{\mathbf{k}\pm\mathbf{q}/2}$ and $\xi_{\mathbf{k}\pm\mathbf{q}/2}$ should be expanded with respect to $q$. It is also crucial to to take into account $k$-dependence of $h_{k}$ near the poles of Eq.(\ref{G}). Some details of such a calculation may be found in Ref.[\onlinecite{Malsh island}]. A lengthy algebra yields the leading term for the quantum correction of the order of  $(q/k_F)(h_{k_F}\tau_{\text{sc}})^3$. In turn, the "electromotive" force $\bm{\mathcal{E}}$ in Eq.(\ref{f2NS}) is given by
\begin{equation}\label{E}
e\bm{\mathcal{E}}=\frac{h_{k_F}^2\tau_{\text{sc}}^2}{2D_N}\bm{\nabla}_{\mathbf{k}}\mathrm{Tr}[\tau_3\mathbf{h}_{\mathbf{k}}\mathbf{g}^K_{N}]\,.
\end{equation}
A similar expression controls the spin-galvanic effect in Josephson junctions \cite{ISHE} and normal systems. \cite{Schwab} In the latter case  $\mathbf{g}^K_{N}$ should be substituted for the spin density.

In addition to the Usadel equations, the nonclassical corrections  also appear in the electric current. The current  is expressed from Eqs.(\ref{Dyson})  in the form
\begin{eqnarray}\label{J}
&& \mathbf{J}(\mathbf{q})=\frac{ie}{4}\int \frac{d\omega}{2\pi}\sum_{\mathbf{k}}\mathrm{Tr}[\mathbf{v}_{\mathbf{k}}\tau_3G^K_{N\mathbf{k}}(\mathbf{q})]=
\nonumber \\
&&\frac{ie}{4}\int \frac{d\omega}{2\pi}\sum_{\mathbf{k}} \mathrm{Tr}[\mathbf{v}_{\mathbf{k}}\tau_3G^r_{N\mathbf{k}+\frac{\mathbf{q}}{2}}\Sigma^K_N(\mathbf{q}) G^a_{N\mathbf{k}+\frac{\mathbf{q}}{2}}]\,,
\end{eqnarray}
where $\mathbf{v}_{\mathbf{k}}=\bm{\nabla}_{\mathbf{k}}(\epsilon_{N\mathbf{k}}+\mathbf{h}_{\mathbf{k}}\bm{\sigma})$. The sought-after nonclassical correction $\mathbf{J}_{\mathrm{nc}}$ can be obtained from the spin dependent part of $\Sigma^K_N(\mathbf{q})$, similar to the above calculation of the correction to the Usadel equation. In the coordinate representation it takes the form
\begin{eqnarray}\label{deltaJ}
\mathbf{J}_{\mathrm{nc}}(\mathbf{r})&=&\frac{e\tau_{\mathrm{sc}}N_{F_N}}{4}\int d\omega\mathrm{Tr}\left[
\Gamma_s  \bm{\nabla}_{\mathbf{k}}(\mathbf{h}_{\mathbf{k}}\tau_3\mathbf{g}^K_{N} )- \nonumber \right.\\
&&\left.2D_N\alpha^2m^*(\mathbf{e}_z\times\bm{\nabla}_{\mathbf{r}})\tau_3g^K_{Nz}\right]\,.
\end{eqnarray}
The first term in the integrand represents the spin-galvanic effect. In the case of Rashba SOC this term gives rise to the electric current in the $x$-direction, if  spins are polarized parallel to the $y$-axis. The corresponding example of the spin injection was considered in Sec.IID. The second contribution to the current stems from spins polarized in the $z$-direction. This is the inverse spin-Hall effect. It has been noted above that in bounded systems, whose size is comparable to $l_{so}$, it is difficult to distinguish these two effects, because $g^K_{Nz}$ and $g^K_{Ny(x)}$ are coupled to each other  via the boundary conditions. In more detail this situation will be analyzed in the next section.

The total current $\mathbf{J}(\mathbf{q})$ consists of $\mathbf{J}_{\mathrm{nc}}$ and usual diffusion currents \cite{Larkin semiclass,Rammer, Kopnin} of quasiparticles in the  normal metal and superconductor, as well as supercurrents due to the order-parameter phase gradient in both layers. We thus obtain for the  current density in the bilayer:
\begin{eqnarray}\label{Jtot}
&&\mathbf{J}=\mathbf{J}_{\mathrm{nc}}+e\int d\omega\left(N_{F_N} \tilde{D}_N^{(2)}\bm{\nabla}f^{(2)}_{0N}+ \nonumber \right.\\ &&\left.d_SN_{F_S}\tilde{D}_S^{(2)}\bm{\nabla}f^{(2)}_{0S}\right)+\left(\frac{en_S}{2m}+\frac{en_N}{2m^*}\right)\bm{\nabla}\chi \,,
\end{eqnarray}
where $n_{S}=2\pi m D_S d_SN_{F_S}\Delta_0\tanh(\Delta_0/2k_BT)$ and $n_{N}=2\pi m^* D_N N_{F_N}\Delta_m\tanh(\Delta_m/2k_BT)$ are 2D densities of superconducting electrons in the normal and superconducting layers. \cite{Kopnin}  Note, that  $N_{F_N}$ is the state density of a 2D gas. The 2D density of states in the superconductor film is given instead by $d_SN_{F_S}$. In Eq.(\ref{Jtot}) the supercurrent in the normal layer may be neglected, because $\Delta_m \ll \Delta_0$. Moreover, one should expect that $d_SN_{F_S}\gg N_{F_N}$, if the superconducting film is thick enough, or the normal system is a 2D electron gas in a semiconductor quantum well. The phase of the order-parameter can be found from the continuity equation $\bm{\nabla}\mathbf{J}=0$. When the operator $\nabla$ is applied to Eq.(\ref{Jtot}) one should take into account Eq.(\ref{f2NS}), Eq.(\ref{tildaT2}) and the relations between the tunneling parameters $T_{NS}$ and $T_{SN}$, that  have been discussed below Eq.(\ref{UsadelS}). Also,  Eqs.(\ref{E}) and (\ref{deltaJ}) give
\begin{equation}\label{Jnc1}
\bm{\nabla} \mathbf{J}_{\mathrm{nc}}=\frac{\sigma_N}{2}\int d\omega\bm{\nabla} \bm{\mathcal{E}}\,,
\end{equation}
where $\sigma_N=2e^2D_NN_{F_N}$ is the conductivity of the normal metal. By this way the equation for the phase can be obtained in the form
\begin{equation}\label{chi}
ed_SN_{F_S}\int d\omega Rf^{(2)}_{0S}-
\frac{en_S}{2m}\nabla^2\chi=0 \,.
\end{equation}

The above equation has been obtained from the charge conservation. It is instructive to derive it in a different way, directly from the gap equation. The latter has the form
\begin{eqnarray}\label{Delta}
&&\frac{\bm{\Delta}}{\lambda}=\frac{1}{8}\int d\omega \mathrm{Tr}_{\tau}[\bm{\tau}g_{0S}^K]=\nonumber\\
&&\frac{1}{8}\int d\omega \mathrm{Tr}_{\tau}[\bm{\tau}(g^r_S-g^a_S)f_{0S}^{(1)}+\bm{\tau}(g^r_S+g^a_S)f_{0S}^{(2)}]\,,
\end{eqnarray}
where $\lambda$ is the electron-electron pairing constant. The first term in the integrand is determined by quasiparticle energies above the gap where, as shown in Sec.IID2, at $k_BT \ll \Delta_0$ and $\mu_s < \Delta_0$ the distribution function $f^{(1)}_{0S}=\tanh(\omega/2k_BT)$, while the retarded and advanced Green functions are given by Eq.(\ref{gras}) (in the rotated representation). The integral of the unperturbed function, that is given by the first term in Eq.(\ref{gras}), cancels with the left-hand side of  Eq.(\ref{Delta}), while the second term in Eq.(\ref{gras}) gives $(1/d_S\Delta_0)(en_S/2m)\nabla^2\chi$. By taking into account Eq.(\ref{D2}) it is easy to see that Eq.(\ref{Delta}) coincides with Eq.(\ref{chi}).
\subsection{Electric current induced by the spin injection}
\subsubsection{Current in a wide strip}
Let us consider  a simple situation of a wide enough strip, such that boundary effects at $y=\pm w/2$ may be neglected. Also, the injected spin polarization will be assumed uniform in the $y$ direction. In the case of Rashba SOC this example was considerd in Sec.IID1 with spins polarized in the $y$-direction. Hence, in Eq.(\ref{deltaJ}) only the first term in the integrand contributes to $\mathbf{J}_{nc}$. From Eq.(\ref{E})  the current density  $J_{nc}$ in the $x$-direction can be expressed in terms of the "electromotive force",
\begin{equation}\label{Jnc}
 J_{nc}=\frac{\sigma_N}{2}\int d\omega \mathcal{E}\,.
\end{equation}
As seen from Eq.(\ref{E}),  Eq.(\ref{fyN}) and Sec.IID1, spatial  variations of $J_{nc}$ are determined by the larger of the Dyakonov-Perel spin-relaxation length $l_{so}$ and the size of the injector. In turn, both lengths have been assumed much smaller than other characteristic lengths of the  system, such as $l_{NS}$, $l_{SN}$ and  $l_R$.
\begin{figure}[tp]
\includegraphics[width=8cm]{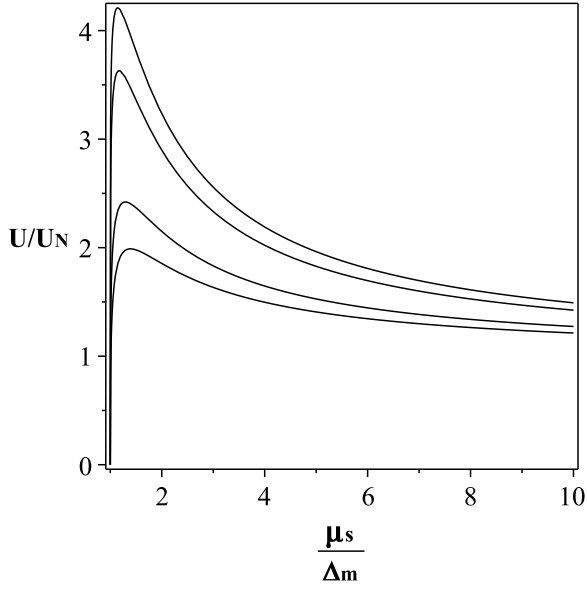}
\caption{The effective electric voltage induced by the spin injection, as a function of the chemical potential difference of two injected spin projections. The voltage is given in units of the effective voltage, that is calculated for a normal metal without the superconducting proximity effect (see text). The chemical potential is measured in units of the minigap in the spectrum of the normal layer. The curves (from top to bottom) are calculated at $2T_{NM}/\Gamma_s=0.005, 0.01, 0.05, 0.1$} \label{fig2}
\end{figure}
Let us consider a solution of Eqs.(\ref{f2NS}) that vanishes at large $x$. This takes place in  a situation when the length of the strip in Fig.1 is larger than all characteristic lengths and an external bias is absent. By substituting this solution into the second term of Eq.(\ref{Jtot}) and combining it with $J_{\mathrm{nc}}$ we obtain the dissipative quasiparticle current $J_d$ in the form
\begin{equation}\label{Jd}
J_d=\frac{\sigma_N}{2}\int d\omega \mathcal{U}\frac{ir_1r_2}{r_1-r_2}\left( \frac{e^{i\sqrt{r_1}|x|}}{\sqrt{r_1}}- \frac{e^{i\sqrt{r_2}|x|}}{\sqrt{r_2}}\right)\,,
\end{equation}
where $\mathcal{U}=\int dx\mathcal{E}$ and $r_1$ and $r_2$ are given by
\begin{eqnarray}\label{r12}
r_1&=&\frac{i}{l_R^2}\frac{\left(l_R^2+il_{NS}^2- l_{NS}^2l_R^2l_{SN}^{-2}\right)}{(l_R^2+il_{NS}^2)}\nonumber \\
r_2&=&-\frac{1}{l_{NS}^2}\frac{\left(l_R^2+il_{NS}^2+l_{NS}^2l_R^2l_{SN}^{-2}\right)}{(l_R^2+il_{NS}^2)}\,.
\end{eqnarray}
In Eq.(\ref{Jd}) the signs of $\sqrt{r_{1(2)}}$  are chosen such, that $\mathrm{Im}(\sqrt{r_{1(2)}})>0$. Since $l_R\ll l_{NS}$ and $ l_{SN}$ we have $r_1 \simeq i/l_R^2$ and  $r_2 \simeq i/l_{NS}^2$, so that $r_1 \gg r_2$. Hence, at a large distance the quasiparticle current is given by the second term in the brackets of Eq.(\ref{Jd}) which, in turn, is determined by the Andreev reflection.

In order to determine the total current, one needs a boundary condition for the order parameter at a large distance. For this, let us assume that the wire has a form of a closed loop having the length $L$.  A change of the phase $\chi$ on this length is $2\pi n$, where $n$ is a whole number. By integrating the current density Eq.(\ref{Jtot}) over the strip area and taking into account that the total current  $I$ is constant and the distribution functions are periodic we obtain
\begin{equation}\label{JL}
 nw\frac{\pi en_S}{m}+\int dxdy J_{nc}=IL\,,
\end{equation}
where the integral of $J_{nc}$ can be obtained from Eq.(\ref{Jnc}). It should be noted that Eq.(\ref{JL}) is valid at an arbitrary relation between $L$ and the relaxation length $l_{NS}$ of the quasiparticle current. If $L \gg l_{NS}$ the most of the current is formed by the condensate, while in the opposite case the current is produced by quasiparticles. In the former case, the second term in Eq.(\ref{JL}) plays the role of an effective magnetic flux through the loop, similar to the equilibrium magnetoelectric effect induced by the Zeeman field.\cite{Malsh island} In the latter case, the electric current has mostly a dissipative nature and it is more reasonable to describe the effect in terms of an effective electromotive force, that is associated with a nonequilibrium spin polarization, as in the case of the spin-galvanic effect in normal metals. The number $n$ in Eq.(\ref{JL}) must be chosen to minimize the energy of the moving condensate, similar to the Little-Parks effect. \cite{Little} Probably, the strong enough spin-galvanic effect might cause a sort of Little-Parks oscillations in the case of $L \gg l_{NS}$. A more realistic possibility, given a weakness of the effect, might be a measuring of a shift in Little-Parks oscillations produced by an external magnetic field. Also, the time-modulated spin injection can significantly affect a flux qubit, when the qubit's resonance frequency coincides with the modulation frequency. In this case the spin injection effect will be similar to an oscillating magnetic flux. \cite{Orlando}

Note, that  in superconducting systems the spin-galvanic effect is very similar to the equilibrium magnetoelectric effect, that is produced by a Zeeman field. In particular, in both cases they result in an effective magnetic flux \cite{Malsh island}. There is, however, a fundamental difference. The Zeeman field gives rise to triplet Cooper pairs, whose dynamics in the Rashba field and conversion to singlet pairs lead to the magnetoelectric effect. In contrast, the spin injection does not produce any changes in the condensate wave function. It modifies  the quasiparticle distribution function only.

The spin injection effect becomes stronger at large SOC. Within the considered theory a strength of this coupling is restricted only by a smallness of the semiclassical parameter $\alpha/v_F$. Also, the effect increases with the larger contact size $b$ of the injector. Therefore, the most interesting case corresponds to $b \gg l_{so}$ in Eq.(\ref{fyN}). At Fig.2 the electromotive voltage $U$, that is defined as $U=\int \mathcal{E}dxd\omega$, is shown as a function of the injection strength $\mu_s$, at various $T_{NM}/\Gamma_s$. In the normal layer at small temperature this voltage is a linear function of $\mu_s$, namely  $U_N=4(b/l_{so})(T_{NM}/\mu_N)\mu_s$. At Fig.2 this value is used as a normalization factor. The nonlinear dependence of $U$ on $\mu_s$  in Fig.2 is associated with the presence of the minigap in the quasiparticle spectrum. By taking $b/l_{so}=5$, $T_{NM}/\mu_N = 10^{-3}$ and $\mu_s=10^{-4} \mathrm{eV} $ we obtain $U \sim U_N \sim 10^{-6}\mathrm{ V}$. The above evaluation of $U$ is mostly restricted by limitations of the theory, which does not allow to take larger $T_{NM}$, $\mu_s$ and $\alpha$. One can not exclude a possibility that larger $U$  may be reached within a more general theory.

\subsubsection{Current in a narrow strip}
As noted in Sec.IID1, in bounded systems  SOC may strongly modify the spatial distribution of the injected spin polarization. In the case of a narrow strip, whose width is less than the spin relaxation length, one must take into account the boundary conditions Eq.(\ref{bc}). The importance of such an analysis becomes evident from Eq.(\ref{deltaJ}) for the nonclassical electric current. Indeed, in the case of Rashba SOC this current can be written in the the form
\begin{equation}\label{Jnc2}
J^x_{\mathrm{nc}}=\frac{e\tau_{\mathrm{sc}}D_{N}N_{N_F}}{2m^*l_{so}^2}\int d\omega
\mathrm{Tr}\left[2\alpha m^*\tau_3g^K_{Ny}+\nabla_y\tau_3g^K_{Nz}\right]\,.
\end{equation}
Since, according to Eq.(\ref{f}) $\mathbf{g}^K\parallel \mathbf{f}$, one can apply the boundary conditions Eq.(\ref{bc}) to the integrand of Eq.(\ref{Jnc2}). As a result, $J^x_{\mathrm{nc}}$  vanishes at the strip boundaries $y=\pm w/2$, because just the expression $2\alpha mf_{Ny}+\nabla_yf_{Nz}$ enters in the integrand of Eq.(\ref{Jnc2}).  Hence, in the case of a narrow strip, whose width $w\ll l_{so}=1/\alpha m^*$, one could expect that $J^x_{\mathrm{nc}}$ is small inside the strip. On the other hand, the Dyakonov-Perel' spin relaxation time increases dramatically in narrow wires.\cite{Malsh wire,Kiselev} Therefore, the spin accumulation in the strip  increases and may compensate a cancellation of the two terms in Eq.(\ref{Jnc2}). In order to check that such a compensation indeed takes place, let us assume that the injector width $b$ is small ($b\ll l_{so}$). Also, in the case of a weak metal-injector coupling one may neglect the back flow of spins from the normal layer. This means that only $\mathbf{f}_M$ must be retained in the tunnel coupling term $\tilde{T}^{(1)}_{NM}(\mathbf{f}_N-\mathbf{f}_M)$ in Eqs.(\ref{spinfN}). A weak coupling to the superconductor will  also be neglected. It is easy to calculate the area integral of $J^x_{\mathrm{nc}}$, that enters in Eq.(\ref{JL}). After integration of Eqs.(\ref{spinfN}) over $x$, the remaining equations may be solved by expanding $\mathbf{f}_{yN}$ and $\mathbf{f}_{zN}$ in power series in $y$, while $\mathbf{f}_{xN}=0$. By this way we obtain
\begin{equation}\label{intxy}
\int dxdy(2\alpha mf_{Ny}+\nabla_yf_{Nz})=-2b\alpha m^*\frac{\tilde{T}^{(1)}_{NM}}{\Gamma_s}f_{yM}\,.
\end{equation}
Let us compare this result with the similar integral obtained in the case of a wide strip. For such a strip $f_{Nz}$ may be neglected, while $f_{Ny}$ is given by Eq.(\ref{fyN}). By expanding this expression with respect to small $b/l_{so}$ we obtain the same result as Eq.(\ref{intxy}). Therefore, there is no difference between wide and narrow wires. On the other hand,  in narrow wires the slower spin relaxation causes an enhanced  leak of the spin polarization into the injector and superconductor.  Consequently, such an effect may become important at larger $b$ and $T_{NM}$.

\section{Conclusion}
It has been shown that the spin-galvanic effect in a hybrid superconductor-Rashba metal bilayer system also has a hybrid character. The injected spin polarization induces both a dissipative quasiparticle current in the normal layer and a supercurrent in the superconducting layer. It depends on the size of the system, which of two effects dominates. There is some characteristic length where a conversion of the quasiparticle's current into the supercurrent through the Andreev reflection occurs. In either of two cases the current of quasiparticles is strongly influenced by the proximity induced minigap in the electron spectrum.

It should be noted that in  bounded systems  one can not observe a pure spin-galvanic effect which is produced by in-plane polarized spins. In such systems Rashba SOC always rotates these spins towards the z-axis. The outplane polarization, in turn, gives rise to the inverse spin-Hall effect. Therefore, there is always a combination of the two effects. A special case of a narrow strip has been considered, whose width is much smaller than the spin precession length in the Rashba field. In such a situation the spin galvanic and the inverse spin-Hall effects tend to cancel each other. Due to the enhanced spin relaxation time in such a narrow wire, the overall effect, however, turned out to be the same as in a wide strip.

Since the spin-galvanic effect in the superconducting condensate can be interpreted as an effective magnetic flux, it adds a new functionality to superconducting quantum circuits and creates a bridge between magnetic and superconducting circuits.


\end{document}